\documentclass[twoside,onecollarge]{article}

\usepackage[applemac]{inputenc}
\usepackage[dvips]{graphicx}
\usepackage{graphicx}
\usepackage[english]{babel}
\usepackage{amsfonts}
\usepackage[reqno]{amsmath}
\usepackage{epsfig}

\usepackage{float}
\floatstyle{boxed}
\restylefloat{figure}

\usepackage{bbm}

\usepackage{color}
\definecolor{lightgray}{gray}{0.85}

\makeindex

\newcommand{\R}{\mathbb{R}}

\newcommand{\E}{\mathbb{E}}

\newcommand{\PP}{\mathbb{P}}

\newcommand{\boA}{\mathcal{A}}

\newcommand{\boL}{\mathcal{L}}

\newcommand{\rnf}{\renewcommand{\thefootnote}{\arabic{footnote}}}
\newcommand{\thankyou}[2]{\stepcounter{footnote}\footnotetext[#1]{#2}}

\title{High-frequency market-making for \\ multi-dimensional Markov processes}

\author{\rnf Pietro Fodra
              \footnotemark[1]
              \footnotemark[2]
\and
\rnf Mauricio Labadie
              \footnotemark[1]
              \footnotemark[3]
}

\date{\today}

\begin{document}
\DeclareGraphicsExtensions{.pdf,.gif,.jpg} \maketitle

\thankyou{1}{\,LPMA (Laboratoire de Probabilit\'es et Mod\`eles Al\'eatoires). Universit\'e Paris Diderot (Paris 7). }

\thankyou{2}{\,EXQIM (Exclusive Quantitative Investment Management). 24 Rue de Caumartin 75009 Paris, France.}

\thankyou{3}{\,Corresponding author. mauricio.labadie@exqim.com, mauricio.labadie@gmail.com}



\maketitle

\begin{abstract}
In this paper we complete and extend our previous work on stochastic control applied to high frequency market-making with inventory constraints and directional bets. Our new model admits several state variables (e.g. market spread, stochastic volatility and intensities of market orders) provided the full system is Markov. The solution of the corresponding HJB equation is exact in the case of zero inventory risk. The inventory risk enters into play in two ways: a path-dependent penalty based on the volatility and a penalty at expiry based on the market spread. We perform perturbation methods on the inventory risk parameter and obtain explicitly the solution and its controls up to first order. We also include transaction costs; we show that the spread of the market-maker is widened to compensate the transaction costs, but the expected gain per traded spread remains constant. We perform several numerical simulations to assess the effect of the parameters on the PNL, showing in particular how the directional bet and the inventory risk change the shape of the PNL density. Finally, we extend our results to the case of multi-aset market-making strategies; we show that the correct notion of inventory risk is the L2-norm of the (multi-dimensional) inventory with respect to the inventory penalties.

\end{abstract}

\bigskip

\noindent {\bf Keywords}: Quantitative Finance, High-Frequency Trading, Market-Making, Inventory Risk, Markov Processes, Hamilton-Jacobi-Bellman, Stochastic Control, Optimal Control.

\maketitle

\tableofcontents

\section{Introduction}

\subsection{Variables}

We will work with time $t\in[0,T]$ and two controls, the half ask (resp. bid) spread of the market-maker $\delta^+$ (resp. $\delta^-$), measured as the distance between the mid-price and her ask quote (resp. bid quote). Our model admits several state variables, which can be any process provided the whole system is Markov. This framework admits a large class of price models e.g. jump processes with stochastic volatility. Without loss of generality, we will restrict our analysis to the folowing ones:
\begin{enumerate}
\item The mid-price $S(t)$, assumed to be an It\^o diffusion.
\item The half market spread $Z(t)$, which is assumed non-negative and somewhat mean-reverting.
\item The volatility $\Sigma(t)$ of the mid-price process, which is strictly positive.
\item The inventory $Q(t)$, which is modelled as
\[
dQ(t) = dN^-(t) - dN^+(t)\,,
\]
where $dN^+(t)$ and $dN^-(t)$ are two independent Poisson processes.
\item The intensities $\lambda^\pm$ of the previous Poisson processes are $(i)$ exponentially decreasing in the distance to the quote on the other side, and $(ii)$ their speed of decay $K$ is random: 
\[
\lambda^\pm(\delta^\pm) = Ae^{-K(t)[z+\delta^\pm]} 
\]

\begin{figure}
\centering
\includegraphics[width=4in]{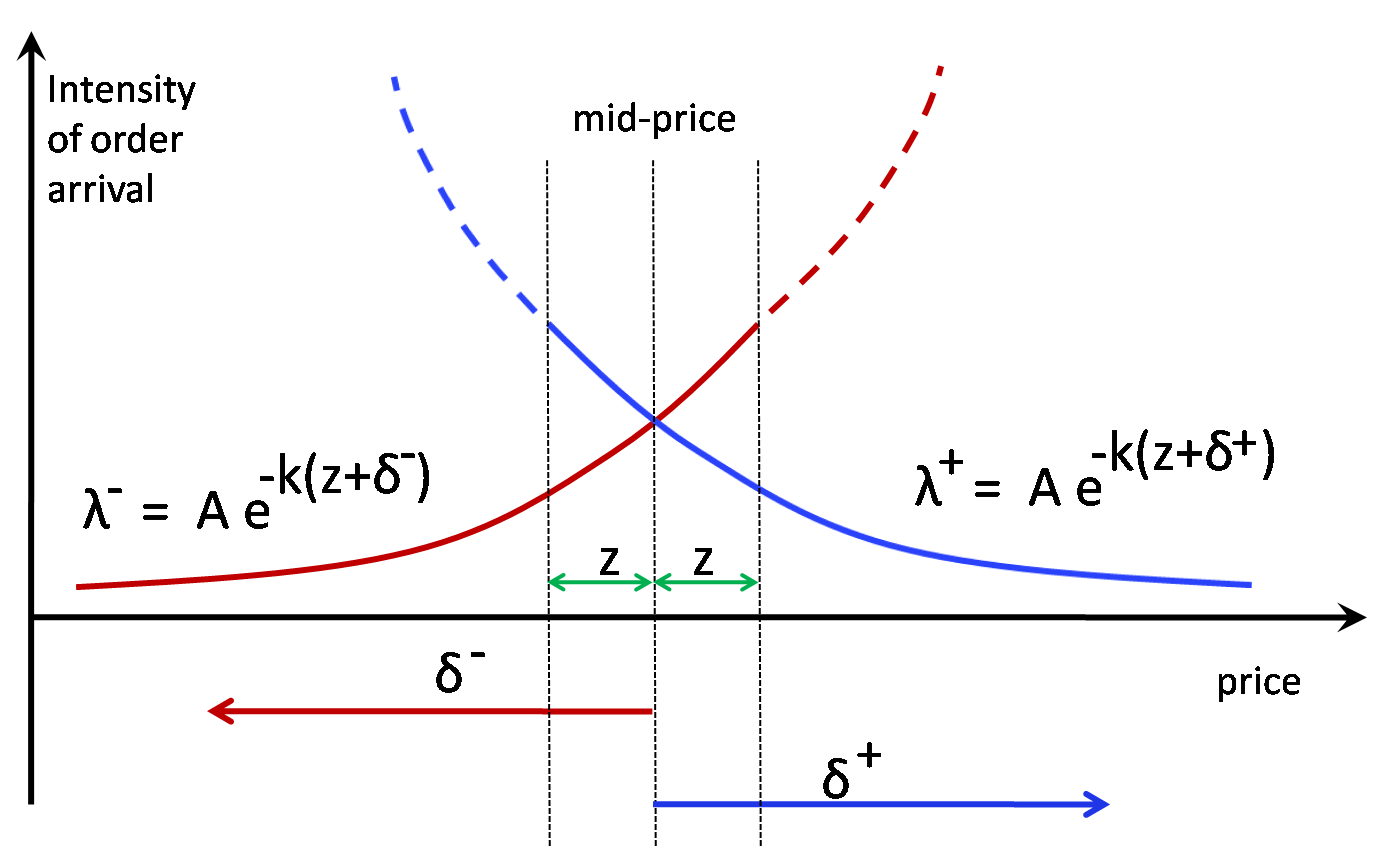}\label{fig-intensity}
\caption{Market order intensities $\lambda^\pm$. They are extrapolated when $\delta^\pm\le -z$ (dotted lines).\label{fig-lambda}}
\end{figure}

\item The cash $X(t)$, which is simply the money earned by the market-making by selling and buying the stock, i.e.
\[
dX(t) = [S(t)+\delta^+]dN^+(t) - [S(t)-\delta^)]dN^-(t)\,.
\]
\end{enumerate}

To keep things simple, we will use the notation
\[
Y(t)=(S(t),Z(t),\Sigma(t),K(t))\quad\text{and}\quad y=(s,z,\sigma,k)\,.
\]
The approach we propose is very general because it admits any number of state variables. In that spirit, one could add to $Y(t)$ other state variables, e.g. a random intensity $A(t)$ of the arrival of market orders or a statistical indicator coming from an alternative market.

\subsection{Hypotheses on the processes and controls}

We assume our controls $\delta^\pm$ lie in an admissible space $\boA$. In order to have a well-posed stochastic control problem, we have to assume that the the system of all state variables is Markov, since otherwise our mathematical techniques do no apply. It is worth to mention that this assumption does not necessarily imply having Markovian variables: as in the case of stochastic volatility, it is not the mid-price which needs to be Markov but the couple (mid-price,volatility).\

Another hypothesis we need is that all value functions $u$ are finite. In the case of the value function \eqref{eq-value-zero}, we accept any mid-price process  $S(t)$ such that the function $u$ in \eqref{eq-sol-zero} is finite. This condition holds for any process such that its conditional expectation $\E_{t,y}[S(T)]$ is affine in $s = S(t)$. For example, a Brownian motion and an Ornstein-Uhlenbeck are acceptable mid-price processes.

\subsection{General HJB and optimal controls}

Let us start with the general HJB equation, i.e.
\begin{eqnarray}\label{eq-hjb-gen}
\left(\partial_t+\boL\right)u + \max_{\delta^+\in\boA}Ae^{-k[z+\delta^+]}\left[u(t,y,q-1,x+(s+\delta^+))-u(t,y,q,x)\right] & & \\
+\max_{\delta^-\in\boA}Ae^{-k[z+\delta^-]}\left[u(t,y,q+1,x-(s-\delta^-))-u(t,y,q,x)\right] &=& - f(t,y,q,x)\,,\nonumber\\
u(T,y,q,x) &=& g(y,q,x))\,,\nonumber
\end{eqnarray}
where $\boL$ is the infinitesimal generator for the state variables $y$. The steps to solve \eqref{eq-hjb-gen} are as follows:
\begin{enumerate}
\item Based on the utility function we make an \emph{ansatz}, i.e. we guess the general form of the solution of \eqref{eq-hjb-gen}, i.e.
\begin{equation}\label{eq-ans}
u(t,y,q,x) = x + v_q(t,s)\,.
\end{equation}

\item We substitute the ansatz \eqref{eq-ans} in \eqref{eq-hjb-gen} in order to find an easier HJB equation for $v$. We use this new HJB equation to find the optimal controls that maximize the jumps. Indeed, after elementary calculus, where the jump part of \eqref{eq-hjb-gen} is considered a function of $(\delta^+,\delta^-)$, one finds that the optimal controls are
\begin{equation}\label{eq-delta-gen}
\delta^+_\ast = \frac{1}{k}-s+v_q(t,y)-v_{q-1}(t,y)\,,\qquad \delta^-_\ast = \frac{1}{k}+s+v_q(t,y)-v_{q+1}(t,y)\,.
\end{equation}

\item We substitute the optimal controls in the HJB equation: the resulting equation is called the \emph{verification equation}. In our case it is
\begin{eqnarray}\label{eq-verif-gen}
\left(\partial_t+\boL\right)v_q + \frac{A}{k}\left(e^{-k[z+\delta^+_\ast]}+e^{-k[z+\delta^-_\ast]}\right) + f(t,y,q,x))&=& 0\,,\\
v_q(T,y) &=& g(y,q,x)\,,\nonumber
\end{eqnarray}
which is highly nonlinear because $\delta^\pm_\ast = \delta^\pm_\ast (v_q)$.

\item We cannot solve directly the verification equation \eqref{eq-verif-gen} via the Feynmann-Kac representation formula because the former is nonlinear and the latter works only for linear equations. However, the idea is to decompose our nonlinear problem into several linear problems, and apply Feynmann-Kac to each one of them. Recall that for a linear equation of the form
\begin{eqnarray*}
\left(\partial_t+\boL\right)w(t,y,q,x) + h(t,y,q,x) &=& 0\,,\\
w(T,y,q,x) &=& g(y,q,x)\,,
\end{eqnarray*}
the (unique) solution is given by the Feynmann-Kac formula (see e.g. Pham \cite{pham}) 
\begin{equation}\label{eq-fk}
w(t,y,q,x) = \E_{t,y,q,x}\left[ g(Y(T),Q(T),X(T)) + \int_t^T h(Y(\xi),Q(\xi),X(\xi))\,d\xi \right]\,,
\end{equation}
where $\E_{t,y,q,x}$ is the conditional expectation given $Y(t-)=y$, $Q(t-)=q$ and $X(t-)=x$. As it will be evident throughout this work, our approach relies entirely on the Feynmann-Kac formula \eqref{eq-fk}. Therefore, as long as our general Markov processes are such that the expectation in \eqref{eq-fk} is finite, we are on solid ground.

\item Alternatively, we could express the optimal quotes in terms of the market-maker's bid-ask spread 
\[
\psi_\ast := \delta^+_\ast+\delta^-_\ast
\]
and the centre of her spread
\[
r_\ast := \frac{1}{2}\left(p^+_\ast+p^-_\ast\right) = s + \frac{1}{2}\left(\delta^+_\ast-\delta^-_\ast\right)\,.
\]
Notice that $\delta^+_\ast=\delta^-_\ast$ if and only if $r_\ast(t)=s=S(t)$. Therefore, $r_\ast-s$ measures  the level of asymmetry of the quotes with respect to the mid-price $s$.
\end{enumerate}

There is an important remark on the shape of the optimal controls \eqref{eq-delta-gen}. If the only state variable is the mid-price then $v_q(t,y)=v_q(t,s)$. Moreover, if the mid-price is a martingale then we can assume that $v_q=v_q(t)$ because, by definition, there is no directional bet on the price. Under these assumptions, plugging the explicit optimal controls \eqref{eq-delta-gen} into the verification equation \eqref{eq-verif-gen} leads to a system of ODEs for $v_q$, indexed by $q$. This system can be solved numerically, but is is nearly impossible to have an explicit formula (see e.g. Cartea-Jaimungal \cite{cartea-risk} and Gu\'eant-Lehalle-Fern\'andez \cite{lehalle}). In our case, with a general price process and several other state variables, there is no hope for an system of ODEs; in fact, the corresponding system is PDE-based. Therefore, either we solve the problem explicitly (as we will do in the linear case), either we perform some asymptotics for the optimal controls (as we will do in the case of inventory penalty).


\section{Linear utility function}

\subsection{Hamilton-Jacobi-Bellman (HJB) equation}

Here we will try to find the optimal controls $\delta^\pm$ that maximise the PNL profit and loss) of a market-maker, i.e. a linear the value function, i.e.
\begin{equation}\label{eq-value-zero}
u(t,y,q,x) = \max_{\delta^\pm\in\boA}\E_{t,y,q,x}
\big[\alpha(Y(T),Q(T),X(T))\big]\,,
\end{equation}
where $\E_{t,y,q,x}$ is the conditional expectation given the values of all variables at time $t$, i.e.
\[
\mathbb{E}_{t,y,q,x}\left[\alpha\right] := \mathbb{E}\left[\alpha\vert Y(t-)=y,Q(t-)=q,X(t-)=x \right]\,.
\]
The probabilistic representation \eqref{eq-value-zero} is the unique solution of the HJB equation
\begin{eqnarray}\label{eq-hjb}
\left(\partial_t+\boL\right)u + \max_{\delta^+\in\boA}Ae^{-k[\delta^++z]}\left[u(t,y,q-1,x+(s+\delta^+))-u(t,y,q,x)\right] & & \\
+\max_{\delta^-\in\boA}Ae^{-k[\delta^-+z]}\left[u(t,y,q+1,z,k,x-(s-\delta^-))-u(t,y,q,x)\right] &=& 0\,,\nonumber\\
u(T,y,q,x) &=& \alpha(s,q,x)\,,\nonumber
\end{eqnarray}
where $\boL$ is the infinitesimal generator for all the continuous state variables $y$. Therefore, we will use \eqref{eq-hjb} to find the optimal controls $\delta^\pm$.

\bigskip

Our linear utility function here is simply the PNL of the market maker, i.e.
\[
\alpha(t,y,q,x) = x + qs\,.
\]
Its corresponding value function is thus
\[
u(t,y,q,x) = \max_{\delta^\pm\in\boA}\mathbb{E}_{t,y,q,x}\left[X(T)+Q(T)S(T) \right]\,.
\]
In this case, the optimal controls are the bid and ask quotes that maximises the PNL of the market-maker throughout the trading day.

\subsection{Ansatz and HJB}

We will look for a solution of the form
\[
u(t,y,q,x) = x + \theta_0(t,y) + q\theta_1(t,y)\,.
\]
With this \emph{ansatz}, the HJB equation takes the form
\begin{eqnarray*}
(\partial_t + \mathcal{L})(\theta_0 + q\theta_1)+\max_{\delta^+\in\boA}Ae^{-k[z+\delta^+]}[s+\delta^+ - \theta_1] + \max_{\delta^-\in\boA}Ae^{-k[z+\delta^-]}[-s+\delta^+ + \theta_1] &=& 0\,,\\
\theta_0(T,y) &=& 0\,,\\
\theta_1(T,y) &=& s\,.
\end{eqnarray*}

\subsection{Computing the optimal controls}

Define
\[
f^+(\delta^+) := Ae^{-k[z+\delta^+]}[s+\delta^+ - \theta_1]\,.
\]
Using elementary Calculus we find that its maximum is attained at
\[
\delta^+_\ast = \frac{1}{k} -s +\theta_1\,.
\]
Analogously, for
\[
f^-(\delta^-) := Ae^{-k[z+\delta^-]}[-s+\delta^-+\theta_1]\,.
\]
the maximum is attained at
\[
\delta^-_\ast = \frac{1}{k} +s -\theta_1\,.
\]
With the optimal half-spreads
\[
\delta^\pm_\ast = \frac{1}{k} \pm (\theta_1 - s)\,,
\]
we can easily compute the optimal spread $\psi_\ast$ for the market maker, along with the centre $r_\ast$ of her spread:
\[
\psi_\ast := \delta^+_\ast + \delta^-_\ast =  \frac{2}{k}\,,\quad r_\ast := s+\frac{1}{2}\left(\delta^+_\ast - \delta^-_\ast\right) = \theta_1\,.
\]
Notice that $\theta_0$ is necessary only for the solution of the HJB equation, not for the controls. Indeed, we only need to find $\theta_1$ in order to have our controls explicit.

\subsection{Solving the verification equation}

With the optimal controls, the HJB equation reduces to
\begin{eqnarray*}
(\partial_t + \mathcal{L})(\theta_0 + q\theta_1)+\frac{A}{k}\left\{e^{-k[z+\delta^+_\ast]} + e^{-k[z+\delta^-_\ast]}\right\} &=& 0\,,\\
\theta_0(T,y) &=& 0\,,\\
\theta_1(T,y) &=& s\,.
\end{eqnarray*}
From the explicit form of $\delta^\pm$ we have
\[
f^+(\delta^+_\ast) := \frac{A}{ek}e^{-kz}e^{-k[\theta_1-s]}\,,\quad f^-(\delta^-_\ast) := \frac{A}{ek}e^{-kz}e^{+k[\theta_1-s]}\,.
\]
Therefore, the verification equation can be rewritten as
\begin{eqnarray}\label{eq-verif-zero}
(\partial_t + \mathcal{L})(\theta_0 + q\theta_1)+\frac{2A}{ek}e^{-kz}\cosh\left(k[\theta_1-s]\right) &=& 0\,,\\
\theta_0(T,y) &=& 0\,,\nonumber\\
\theta_1(T,y) &=& s\,.\nonumber
\end{eqnarray}
We separate \eqref{eq-verif-zero} into two equations, one for each one of our unknowns:
\begin{eqnarray*}
(\partial_t + \mathcal{L})\theta_0+\frac{2A}{ek}e^{-kz}\cosh\left(k[\theta_1-s]\right) &=& 0\,,\\
\theta_0(T,y) &=& 0\,,\\
\end{eqnarray*}
and
\begin{eqnarray*}
(\partial_t + \mathcal{L})\theta_1&=& 0\,,\\
\theta_1(T,y) &=& s\,.
\end{eqnarray*}
Let us define
\[
\Delta(t,y) := \mathbb{E}_{t,y}\left[S(T)\right] - s\,,
\]
which measures the difference between the expected value of the mid-price at maturity and its current value. With this notation, and using the Feynman-Kac formula twice, first for $\theta_1$ and then for $\theta_1$, we find
\begin{eqnarray*}
\theta_1(t,y)  &=& s +\Delta(t,y)\,,\\
\theta_0(t,y)  &=& \frac{2}{e}\mathbb{E}_{t,y}\left[\int_t^T\frac{A}{K}e^{-KZ}\cosh(K\Delta)\,d\xi\right]\,,
\end{eqnarray*}
where all capital letters inside the integral are evaluated at $\xi$, i.e. $K=K(\xi)$, $\Delta=\Delta(\xi,Y(\xi))$, etc. This leads to explicit expressions for the (unique) solution and its controls:
\begin{eqnarray}\label{eq-sol-zero}
u(t,y,q,x) &=& u_{\text{hold}}(t,y,q,x) + u_{\text{mm}}(t,y)\,,\\
u_{\text{hold}} &=& x + q(s + \Delta)\,,\nonumber\\
u_{\text{mm}} &=&\frac{2}{e}\mathbb{E}_{t,y}\left[\int_t^T\frac{A}{K}e^{-KZ}\cosh(K\Delta)\,d\xi\right]\,,\nonumber\\
\delta^\pm_\ast &=& \frac{1}{k} \pm \Delta\,,\nonumber\\
\psi_\ast &=& \frac{2}{k}\,,\nonumber\\
r_\ast &=&  s +\Delta\,.\nonumber
\end{eqnarray}

\subsection{Remarks}

Let us explain the \textbf{decomposition of the solution $u$ given in \eqref{eq-sol-zero} in terms of $u_{\text{hold}}$ and $u_{\text{mm}}$}. On the one hand, the function $u_{\text{hold}}$ is the expectation at expiry $T$ of the current portfolio $x+qs$; this corresponds to a \emph{buy-and-hold} strategy. On the other hand, the function $u_{\text{mm}}$ (as the integral from $t$ to $T$ suggests) is the profit for playing a dynamic (i.e. high-frequency) market-making strategy. Since $u_{\text{mm}}>0$, the addition of the market-making mechanism $u_{\text{mm}}$ to the strategy $u$ is more profitable than the buy-and-hold strategy $u_{\text{hold}}$ alone. In consequence, it makes sense to play market-maker dynamically instead of simply apply a buy-and-hold strategy.\

Since in principe $T$ can be very big, in order to compare strategies we have to compute the value function per time unit, i.e. $u(t,y,q,x)/(T-t)$. Using the integral version of the mean-value theorem, it follows that there exists in $\tau\in(t,T)$ such that
\[
\frac{u(\tau,y,q,x)}{T-t} =  \frac{x + q(s + \Delta)}{T-t} + \frac{2}{e}\mathbb{E}_{\tau,y}\left[\frac{A}{K}e^{-KZ}\cosh(K\Delta)\right]\,.
\]
In particular, when $t=0$ we have $q=0$, $x=0$ and $\Delta=0$. Therefore, at the beginning of the day the expected gain of the market-maker is
\[
\frac{u(0,y,0,0)}{T} =  \frac{2}{e}\mathbb{E}_{\tau,y}\left[\frac{A}{K}e^{-KZ}\cosh(K\Delta)\right]\ge \frac{2}{e}\mathbb{E}_{\tau,y}\left[\frac{A}{K}e^{-KZ}\right]\,.
\]
The first observation is that \textbf{the worst mid-price dynamic is the martingale}. Indeed, $\xi\mapsto\cosh(\xi)$ has a strict minimum when $\xi=0$, and a non-martigale process has times $\tau$ where $\Delta(\tau)\neq0$. This seems counter-intuitive because, the market-making being lightning fast, it should not be influenced by the long-range behaviour of the mid-price. However, the inventory turnover is slower because it takes several trades to build up and come back to zero, and that is where the directional bet $\cosh(K\Delta)$ enters into the game.

Another feature is \textbf{the effect of a non-constant decay rate $K$}. Notice that the optimal spread $\psi_\ast$ is decreasing in $k$ and the value function $u$ is decreasing in $K$. If the intensity of order arrival increases, which can be interpreted as either less market orders or a more populated limit order book (LOB), then the market-maker has to reduce her spread to keep her order flow constant. But this makes her PNL smaller because she is selling liquidity cheaper.\

\textbf{The effect of the market spread $Z$} is very interesting. On the one hand, $z$ does not appear at all in the optimal controls, which means that the market-making strategy is independent of $z$. On the other hand, the value function $u$ is decreasing in $Z$. In consequence, the PNL of the strategy decreases as the spread increases. This is a direct consequence of the hypothesis that the intensity of the arrival of market orders depends on the distance to the other side of the book, not on the distance to the mid-price. Indeed, the bigger the spread, the less market flow captured by the market-maker, even if her position relative to the mid-price does not change.\

We have put the intensity $A$ into the expectation operator. This is because our formulation allows a \textbf{stochastic intensity $A(t)$}. In fact, our formulation allows as well \textbf{asymmetric intensities $A^\pm(t)$ and decays $K^\pm(t)$}; the only difference is that the symmetry via $\cosh(\cdot)$ is lost and we would have 2 different, complicated exponentials instead.


\section{Linear utility function with inventory penalty}

\subsection{Two inventory penalties}

We will penalise the inventory in two ways:
\begin{enumerate}
\item A penalty $\Pi_1$ at expiry, depending on the spread. This models the fact that the market-maker will have to clear her inventory at the market, and as such she will pay the spread for each share:
\[
\Pi_1(T)=\eta Z(T)Q^2(T)\,,\quad \eta\ge0\,.
\]
For example, if $z\equiv 1$ we recover the Stoll model for a quadratic penalty on the inventory (see \cite{stoll}).

\item An integral penalty $\Pi_2$ of the (squared) inventory during the remaining of the trading session, weighted by the volatility $\Sigma(t)$ of the mid-price $S(t)$. This is a sort of tracking error with respect to a flat-inventory position, and is a very standard choice (see e.g. Guilbaud-Pham \cite{pham-hft} and Cartea-Jaimungal \cite{cartea-risk}):
\[
\Pi_2(T)=\nu\int_t^T \Sigma^2(\xi)Q^2(\xi)\,d\xi\,,\quad \nu\ge0\,.
\]
\end{enumerate}
In this section, we will consider the following value function:
\begin{equation}\label{eq-u-z}
u(t,y,q,x) = \max_{\delta^\pm\in\boA}\E_{t,y,q,x}\Bigl[X(T)+Q(T)S(T)-\varepsilon \Pi(T)\Bigr]\,,\quad \Pi := \Pi_1 +\Pi_2\,.
\end{equation}
Notice that if $\varepsilon=0$ we recover the previous linear case. Concerning the parameters, the important one is $\varepsilon$ since it gives the penalty as a perturbation of the value function we already computed. The other parameters $\eta$ and $\nu$ can be considered as booleans, so that we can assess \emph{a porteriori} the effect of the two penalties on the resulting controls.

\subsection{The HJB equation and the ansatz}

The resulting HJB equation is thus
\begin{eqnarray}\label{eq-hjb-epsi}
\left(\partial_t+\boL\right)u + \max_{\delta^+\in\boA}Ae^{-k[z+\delta^+]}\left[u(t,y,q-1,x+(s+\delta^+))-u(t,y,q,x)\right] & & \\
+\max_{\delta^-\in\boA}Ae^{-k[z+\delta^-]}\left[u(t,y,q+1,x-(s-\delta^-))-u(t,y,q,x)\right] &=& \varepsilon\nu\sigma^2 q^2\,,\nonumber\\
u(T,y,q,x) &=& x+sq-\varepsilon\eta zq^2\,.\nonumber
\end{eqnarray}
Recall that we have an explicit formula for the (unique) solution when $\varepsilon=0$. Therefore, we will use perturbation methods on $\varepsilon$. More precisely, we propose the following \emph{ansatz}:
\begin{eqnarray}\label{eq-ans-z}
u(t,y,q,x) &=& x + v^{(0)}_q(t,y) + \varepsilon v^{(1)}_q(t,y) + O\left(\varepsilon^2\right)\,,\\
v^{(0)}_q(t,y) &=& \theta^{(0)}_0(t,y) + q\theta^{(0)}_1(t,y)\,,\nonumber\\
v^{(1)}_q(t,y) &=& \theta^{(1)}_0(t,y) + q\theta^{(1)}_1(t,y) + q^2\theta^{(1)}_2(t,y)\,.\nonumber
\end{eqnarray}

\subsection{Verification equation and its linearisation}

With the ansatz \eqref{eq-ans-z}, the optimal controls take the form
\begin{eqnarray}\label{eq-delta-z}
\delta^+_\ast &=& \frac{1}{k}-s+v^{(0)}_q-v^{(0)}_{q-1} + \varepsilon\left[v^{(1)}_q-v^{(1)}_{q-1} \right] + O\left(\varepsilon^2\right)\,,\\
\delta^-_\ast &=& \frac{1}{k}+s+v^{(0)}_q-v^{(0)}_{q+1} + \varepsilon\left[v^{(1)}_q-v^{(1)}_{q+1} \right] + O\left(\varepsilon^2\right)\,.\nonumber
\end{eqnarray}
Under these conditions, the verification equation is
\begin{eqnarray}\label{eq-verif-nonlin1}
\left(\partial_t+\boL\right)\left(v^{(0)}_q + \varepsilon v^{(1)}_q\right) &+& \frac{A}{ek}e^{-kz}e^{ks}\exp\left\{-k\left(v^{(0)}_q-v^{(0)}_{q-1} + \varepsilon\left[v^{(1)}_q-v^{(1)}_{q-1} \right]\right) + O(\varepsilon^2)\right\} \\
&+& \frac{A}{ek}e^{-kz}e^{-ks}\exp\left\{-k\left(v^{(0)}_q-v^{(0)}_{q+1} + \varepsilon\left[v^{(1)}_q-v^{(1)}_{q+1} \right]\right) + O(\varepsilon^2)\right\}\nonumber\\
&=& \varepsilon\nu\sigma^2 q^2\,,\nonumber\\
v^{(0)}_q(T,y) &=& sq\,,\nonumber\\
v^{(1)}_q(T,y) &=& -\varepsilon\eta zq^2\,.\nonumber
\end{eqnarray}
Observe that we can rewrite the jump term in $\delta^+_\ast$
as
\[\frac{A}{ek}e^{-kz}e^{ks}e^{-k\theta^{(0)}_1} \exp\left\{ -k\varepsilon\left[v^{(1)}_q-v^{(1)}_{q-1} \right] + O(\varepsilon^2)\right\}\,.
\]
In consequence, its first-order expansion in $\varepsilon$ is 
\[
\frac{A}{ek}e^{-kz}e^{-k[\theta^{(0)}_1-s]} \left\{1 -k \varepsilon\left[v^{(1)}_q-v^{(1)}_{q-1} \right] + O\left(\varepsilon^2\right)\right\}\,.
\]
Analogously, for the term in $\delta^-_\ast$ we have
\[
\frac{A}{ek}e^{-kz}e^{k[\theta^{(0)}_1-s]}\times \left\{1 -k \varepsilon\left[v^{(1)}_q-v^{(1)}_{q+1} \right] + O\left(\varepsilon^2\right)\right\}\,.
\]
Therefore, it follows that the linearisation in $\varepsilon$ of the verification equation \eqref{eq-verif-nonlin1} is
\begin{eqnarray}\label{eq-verif-lin1}
\left(\partial_t+\boL\right)\left(v^{(0)}_q + \varepsilon v^{(1)}_q\right) &+& \frac{A}{ek}e^{-kz}e^{-k[\theta^{(0)}_1-s]} \left\{1 -k  \varepsilon\left[v^{(1)}_q-v^{(1)}_{q-1} \right]\right\} \\
&+& \frac{A}{ek}e^{-kz}e^{+k[\theta^{(0)}_1-s]} \left\{1 -k  \varepsilon\left[v^{(1)}_q-v^{(1)}_{q+1} \right]\right\}\nonumber\\
&=& \varepsilon\nu\sigma^2 q^2\,,\nonumber\\
v^{(0)}_q(T,y) &=& sq\,,\nonumber\\
v^{(1)}_q(T,y) &=& -\varepsilon\eta zq^2\,.\nonumber
\end{eqnarray}

\subsection{Solution for $\varepsilon^0$}

At zero-th order in $\varepsilon$, which is equivalent to set $\varepsilon=0$, we recover the verification equation of the linear case without inventory penalty \eqref{eq-verif-zero}, i.e.
\begin{eqnarray}\label{eq-verif-epsi0}
\left(\partial_t+\boL\right)\left(\theta^{(0)}_0 + q\theta^{(0)}_1 \right) + \frac{2A}{ek}e^{-kz}\cosh\left(k[\theta^{(0)}_1-s]\right) &=& 0\,,\nonumber\\
\theta^{(0)}_0(T,y) &=& 0\,,\nonumber\\
\theta^{(0)}_1(T,y) &=& s\,.\nonumber
\end{eqnarray}
Therefore, $v^{(0)}_q(t,y)$ is exactly the solution we found before, i.e.
\begin{eqnarray}\label{eq-theta-epsi0}
v^{(0)}_q(t,y) &=& \theta^{(0)}_0(t,y) + q\theta^{(0)}_1(t,y) \,,\\
\theta^{(0)}_1(t,y)  &=& s+\Delta\,,\nonumber\\
\theta^{(0)}_0(t,y)  &=& \frac{2}{e}\mathbb{E}_{t,y}\left[\int_t^T\frac{A}{K}e^{KZ}\cosh(K\Delta)\,d\xi\right]\,,\nonumber
\end{eqnarray}
where (as before) $\Delta := \mathbb{E}_{t,y}\left[S(T)\right] - s$. This fact is our main motivation for the application of perturbation methods on the inventory constraints.

\subsection{Solution for $\varepsilon^1$}

Keeping only the first-order terms in \eqref{eq-verif-lin1}, i.e. those with $\varepsilon$, we obtain
\begin{eqnarray}\label{eq-verif-epsi1}
\left(\partial_t+\boL\right)\left(\theta^{(1)}_0 + q\theta^{(1)}_1 + q^2\theta^{(1)}_2\right) &+& \frac{A}{e}e^{-kz}e^{-k\Delta} \left\{-\theta^{(1)}_1 + (1-2q)\theta^{(1)}_2\right\} \\
&+& \frac{A}{e}e^{-kz}e^{+k\Delta} \left\{\theta^{(1)}_1 + (1+2q)\theta^{(1)}_2\right\}\nonumber\\
&=& \nu\sigma^2 q^2\,,\nonumber\\
\theta^{(1)}_0(T,y) &=& 0\,,\nonumber\\
\theta^{(1)}_1(T,y) &=& 0\,,\nonumber\\
\theta^{(1)}_2(T,y) &=& -\eta z\,.\nonumber
\end{eqnarray}
We decompose \eqref{eq-verif-epsi1} into three equations, and as before we solve one by one via the Feynmann-Kac formula. The first equation, which regroups the terms in $q^2$, is linear:
\begin{eqnarray*}
\left(\partial_t+\boL\right)\theta^{(1)}_2 &=& \nu\sigma^2\,,\\
\theta^{(1)}_2(T,y) &=& -\eta z\,.
\end{eqnarray*}
Its (unique) solution is
\begin{equation}\label{eq-theta-12}
\theta^{(1)}_2(t,y)=-\eta\E_{t,y}\left[Z(T)\right]-
\nu\E_{t,y}\left[\int_t^T\Sigma^2\,d\xi\right]\,.
\end{equation}
Since $\theta^{(0)}_1$ and $\theta^{(1)}_2$ are already known, the equation for the terms $q^1$ becomes linear in $\theta^{(1)}_1$, i.e.
\begin{eqnarray*}
\left(\partial_t+\boL\right)\theta^{(1)}_1 + \frac{4A}{e}e^{-kz}\sinh(k\Delta)\theta^{(1)}_2&=& 0\,.
\\
\theta^{(1)}_1(T,y) &=& 0\,.
\end{eqnarray*}
Therefore, its (unique) solution is
\begin{equation}\label{eq-theta-11}
\theta^{(1)}_1(t,y) = \frac{4}{e}\E_{t,y}\left[\int_t^T Ae^{-KZ}\sinh(K\Delta)\theta^{(1)}_2\,d\xi\right]\,.
\end{equation}
Finally, for $q^0$ we get
\begin{eqnarray*}
\left(\partial_t+\boL\right)\theta^{(1)}_0 &+& \frac{A}{e}e^{-kz}e^{-k\Delta} \left\{-\theta^{(1)}_1 + \theta^{(1)}_2\right\} \\
&+& \frac{A}{e}e^{-kz}e^{+k\Delta} \left\{\theta^{(1)}_1 + \theta^{(1)}_2\right\} = 0\,,\\
\theta^{(1)}_0(T,y) &=& 0\,,
\end{eqnarray*}
whose (unique) solution is
\begin{equation}\label{eq-theta-10}
\theta^{(1)}_0(t,y) = \frac{2}{e}\E_{t,y}\left[\int_t^T Ae^{-KZ}\left\{\theta^{(1)}_1\sinh(k\Delta)+
\theta^{(1)}_2\cosh(k\Delta)\right\}\,d\xi\right]\,.
\end{equation}

Putting all together \eqref{eq-ans-z}, \eqref{eq-theta-epsi0}, \eqref{eq-theta-12}, \eqref{eq-theta-11}, \eqref{eq-theta-10} we finally obtain the explicit expansion of $u(t,y,q,x)$ up to first order in $\varepsilon$.

\subsection{Optimal controls}

Now we are in measure to write down explicitly the optimal controls up to first-order in $\varepsilon$. From the ansatz \eqref{eq-ans-z} and the control equations \eqref{eq-delta-z} it follows that
\begin{eqnarray}\label{eq-delta-theta}
\delta^+_\ast &=& \frac{1}{k}-s+\theta^{(0)}_1+\varepsilon\Bigl(\theta^{(1)}_1 - (1-2q)\theta^{(1)}_2 \Bigr)+O\left(\varepsilon^2\right)\,,\\
\delta^-_\ast &=& \frac{1}{k}+s-\theta^{(0)}_1+\varepsilon\Bigl(-\theta^{(1)}_1 - (1+2q)\theta^{(1)}_2 \Bigr)+O\left(\varepsilon^2\right)\,.\nonumber
\end{eqnarray}
This implies that the optimal spread for the market-maker is
\begin{equation}\label{eq-psi-theta}
\psi_\ast = \delta^+_\ast+\delta^-_\ast = \frac{2}{k}-2\varepsilon\theta^{(1)}_2+O\left(\varepsilon^2\right)\,,
\end{equation}
whilst the centre of her spread is
\begin{equation}\label{eq-r-theta}
r_\ast = s + \frac{1}{2}\left(\delta^+_\ast-\delta^-_\ast\right) = \theta^{(0)}_1 + \varepsilon\Bigl(\theta^{(1)}_1 +2q\theta^{(1)}_2\Bigr)+ O\left(\varepsilon^2\right)\,.
\end{equation}
In the light of all the former computations, we can now write down in full splendour the optimal controls for the market-maker:
\begin{eqnarray}\label{eq-controls-epsi}
\delta^+_\ast &=& \frac{1}{k} + \Delta +\varepsilon\left\{-\E_{t,y}\left[\int_t^T H\tilde\pi d\xi\right]+(1-2q)\tilde\pi\right\}+
O\left(\varepsilon^2\right)\,,\\
\delta^-_\ast &=& \frac{1}{k} - \Delta +\varepsilon\left\{\E_{t,y}\left[\int_t^T H\tilde\pi d\xi\right]+(1+ 2q)\tilde\pi\right\}+
O\left(\varepsilon^2\right)\,,\nonumber\\
\psi_\ast &=&  \frac{2}{k}+2\varepsilon\tilde\pi
+O\left(\varepsilon^2\right)\,,\nonumber\\
r_\ast &=& s+\Delta - \varepsilon\left\{ \E_{t,y}\left[\int_t^T H\tilde\pi d\xi\right]+2q\tilde\pi\right\}+
O\left(\varepsilon^2\right)\,,\nonumber
\end{eqnarray}
where $\Delta := \mathbb{E}_{t,y}\left[S(T)\right] - s$ is the degree of non-martingality of the mid-price (i.e. the directional bet), 
\[
H:= \frac{4}{e} Ae^{-KZ}\sinh(K\Delta)
\]
is the (marginal) profit of the market-making as a fuction of the directional bet $\Delta$, which is a function of $\Delta$, and
\[
\tilde\pi := \eta\E_{t,y}\left[Z(T)\right]+
\nu\E_{t,y}\left[\int_t^T\Sigma^2\,d\xi\right]>0
\]
is the unitary inventory-risk penalty.

\subsection{Remarks}

\textbf{The unitary inventory-risk penalty $\tilde\pi$} has two components. The term in $\nu$ (the integral one) penalises a non-flat inventory via the volatility thoughout the day. The term in $\eta$ is a penalty that triggers only near the end of the day, but strong enough to force the market-maker to leave the trading floor with zero inventory. Both penalties are complementary: one keeps the inventory within range during the trading day whilst the other forces the inventory to finish the day near zero.

Notice that \textbf{the optimal spread $\psi_\ast$ is increasing in the unitary inventory-risk penalty $\tilde\pi$}. This feature can be understood in a very intuitive way: if a market-maker is more sensitive to an inventory risk then she will be more conservative in her quotes, fearing that even a small price jump would put her on the wrong side of the trend. Her inventory-risk aversion thus translates into a wider spread. However, the inventory $q$ does not appear at all in the expression for $\psi_\ast$. Indeed, it is the \emph{unitary} inventory-risk aversion which determines the width of the spread, not the current inventory level.
 
Observe that \textbf{the center $r_\ast$ of the spread is decreasing in the inventory $q$}. This is also very intuitive: the market-maker will tilt her quotes, rendering them asymmetrical, in order to favour execution vs incoming market orders which help her reduce her (absolute) inventory. For example, if she is long inventory ($q>0$) then she will post aggressive bid quotes to lure selling market orders. At the same time, her ask quotes are more conservative because she does not want to increase her inventory via buying market orders. If $q<0$ then the market-maker will post conservative sell orders and aggressive buy orders, hoping to reduce her inventory via an asymmetrical market-flow.\

\textbf{The directional bet $\Delta$ is present in the centre $r_\ast$}. For example, if $\Delta>0$ then the market-maker expects a final mid-price higher than the current one. Therefore, if she is willing to carry some inventory-risk, she will post aggressive bid quotes and less aggressive ask quotes. As a result, in average the inventory will be positive, reflecting the long bet in the mid-price. As we can see, the dynamic of the centre $r_\ast$ is governed by two opposite effects: $\Delta$ tries to build up the inventory to profit from the difference between the current mid-price and the estimate of the final mid-price, whilst the term in $\varepsilon$ aims to keep a flat-inventory position.\

If the midprice is a martingale then $\Delta\equiv0$ and $r_\ast = s -2q\varepsilon\tilde\pi + O(\varepsilon^2)$, i.e. we get rid of the integral term in $\sinh(\cdot)$. This implies that \textbf{the integral term in $\sinh(\cdot)$ measures the profit of the market-making strategy with respect to the directional bet $\Delta\neq0$}. Now assume that in the non-martingale case we want more simplicity on the formulas, namely we totally discard the integral terms in $\sinh(\cdot)$ and in the volatility $\Sigma^2$; in fact, this is equivalent of a zero-th order expansion in $T-t$ of the optimal controls. It turns out that we recover the optimal quotes in Fodra and Labadie \cite{fodra-labadie}, which were obtained by linearising the verification equation without much care on the accuracy of the approximation. This means that the integral terms, which are are a path-dependent, offer a correction based on the time to maturity $T-t$ and the degree of non-martingality $\Delta$.

\subsection{The effect of transaction costs}

Suppose that the market-maker pays a fixed fee of $\alpha$ for each traded asset. In most venues $\alpha>0$, but there are some trading platforms where a liquidity provider receives a rebate, i.e. $\alpha<0$. Since the transaction cost $\alpha$ affects the cash process each time a share is traded, we have to modify \eqref{eq-hjb-gen} as
\begin{eqnarray}\label{eq-hjb-cost}
\left(\partial_t+\boL\right)u + \max_{\delta^+\in\boA}Ae^{-k[z+\delta^+]}\left[u(t,y,q-1,x+(s+\delta^+)-\alpha)-u(t,y,q,x)\right] & & \\
+\max_{\delta^-\in\boA}Ae^{-k[z+\delta^-]}\left[u(t,y,q+1,x-(s-\delta^-)-\alpha)-u(t,y,q,x)\right] &=& \varepsilon\nu\sigma^2 q^2\,,\nonumber\\
u(T,y,q,x) &=& x+sq-\varepsilon\eta zq^2\,.\nonumber
\end{eqnarray}

Let $\delta^\pm_\ast$ be the optimal controls without transaction costs, and $\delta^\pm_{\alpha\ast}(z)$ the optimal controls under transaction costs. If we define
\[
H(\alpha):= \frac{4}{e}Ae^{-K(Z+\alpha)}\sinh(K\Delta)
\]
then repeating the previous computations we obtain
\begin{eqnarray}\label{eq-controls-alpha}
\delta^+_{\alpha\ast} &=& \frac{1}{k}  + \alpha + \Delta +\varepsilon\left\{-\E_{t,y}\left[\int_t^T H(\alpha)\tilde\pi d\xi\right]+(1-2q)\tilde\pi\right\}+
O\left(\varepsilon^2\right)\,,\\
\delta^-_{\alpha\ast} &=& \frac{1}{k}  + \alpha - \Delta +\varepsilon\left\{\E_{t,y}\left[\int_t^T H(\alpha)\tilde\pi d\xi\right]+(1+ 2q)\tilde\pi\right\}+
O\left(\varepsilon^2\right)\,,\nonumber\\
\psi_{\alpha\ast} &=&  \frac{2}{k} + 2\alpha +2\varepsilon\tilde\pi
+O\left(\varepsilon^2\right)\,,\nonumber\\
r_{\alpha\ast} &=& s+\Delta - \varepsilon\left\{ \E_{t,y}\left[\int_t^T H(\alpha)\tilde\pi d\xi\right]+2q\tilde\pi\right\}+
O\left(\varepsilon^2\right)\,.\nonumber
\end{eqnarray}

As we can see, the transaction cost $\alpha$ widens the market-maker's spread by $2\alpha$, i.e.
\[
\psi_{\alpha\ast} = \psi_\ast + 2\alpha\,.
\] 
If we recall that each time the market-maker trades the spread she pays exactly $2\alpha$, we conclude that her gain per traded spread is constant with value $\psi_\ast$, regardless of the transaction cost $\alpha$. This feature has two immediate consequences:
\begin{itemize}
\item If $\alpha>0$ then the market-maker tries to compensate her loss in transaction costs by widening her spread by $2\alpha$. By acting this way, she keeps her gain per traded spread constant but accepts a smaller probability of execution since she is moving her quotes further into the LOB. If $\alpha$ is too big then her spread could be so wide and far from the best quotes that the market-maker will not capture any market flow at all. Now imagine a scenario where most market-makers have the same behaviour: if they all have wider spreads then this would in aggregate increase the market spread $2z$. In consequence, the liquidity would diminish since the spread gets bigger and thus market orders will get executed at a price further from the mid-price.

\item In the presence of a rebate (i.e. $\alpha<0$) the market-maker systematically reduces her optimal spread. More precisely, she uses the rebate to improve her quotes, keeping her gain per traded spread constant but increasing her probability of execution by placing bets closer to the mid-price. If $\alpha$ is negative enough to have $\psi_{\alpha\ast}=0$ then the market-maker could even accept to buy and sell at the same price, earning no profit in the operation except for the rebate. This strategy is called \emph{scalping} or \emph{rebate arbitrage}, and has been commented by several practitioners (see e.g. Bodek and Shaw \cite{bodek-scalping}). 
\end{itemize} 

From the previous observations, it follows that the fee structure in high-frequency trading is crucial for the liquidity offer (measured as both the size of the spread and the available volume in the LOB) and the profitability of high-frequency market-making strategies. Moreover, trading venues with a rebate mechanism have smaller spreads, are more profitable for market-makers and lure non-market-making arbitrage strategies. These findings, although they are based on our model and its hypotheses, are consistent with several other theoretical and empirical studies on high-frequency trading.\

For example, Colliard and Hoffmann \cite{colliard} found that the French financial transaction tax (FTT) imposed on August 2012 has reduced the number of aggressive limit orders, i.e. those that reduce the spread. The FTT also reduced the depth of the market, i.e. the sum of available volume at best bid and best ask, as well as the trading volume. Finally, the market spread and the volatility were unaffected.\

Comparing our results with those of Colliard and Hoffman, one could argue that the loss of volume at the best quotes and inside the market spread was due to traders increasing their spreads as a response to an increase of $\alpha$. That said, it is worth to mention that FTT did not apply to market-makers who are members of the exchanges. In consequence, these fee-exempted market-makers continued offering liquidity at the best quotes, which explains why the market spread did not change with the FTT implementation.


\section{Numerical Simulations}

We performed several numerical simulation in order to assess quantitatively the statistical properties of the market-making strategy and its parameters, esp. the mid-price pattern (martingale or mean-reverting), the inventory risk $\varepsilon$ and the transaction costs $\alpha$.  All simulations were performed in R.\\

The mid-price in all cases will be an Ornstein-Uhlenbeck process, i.e.
\[
dS(t) = a[S(t)-\mu] + \sigma dW(t)\,,
\]
which satisfies
\begin{eqnarray*}
\E_{t,s}[S(t)] &=& se^{-a(T-t)} + \mu(1-e^{-a(T-t)})\,\\
\Delta &=& (\mu - s)\left(1-e^{-a(T-t)}\right)\,.
\end{eqnarray*}
Recall that for a martingale (e.g. Brownian motion) we have $\E_{t,s}[S(t)]=s$ and $\Delta=0$.\\

The parameters we use, unless explicitly stated in the figures, are 10,000 simulations, 1000 events per day, $T=1$, $A=1000$, $k=1$, initial price $S(0)=3000$, $a=0.1$, $\mu=3009$ (i.e. +3\% of initial price), $\sigma = 0.5$, $z=0.5$, $\varepsilon = 0.001$, $\eta=1$, $\nu=1$, $\alpha=0.05$.\\

\begin{figure}
\centering
\includegraphics[width=5in]{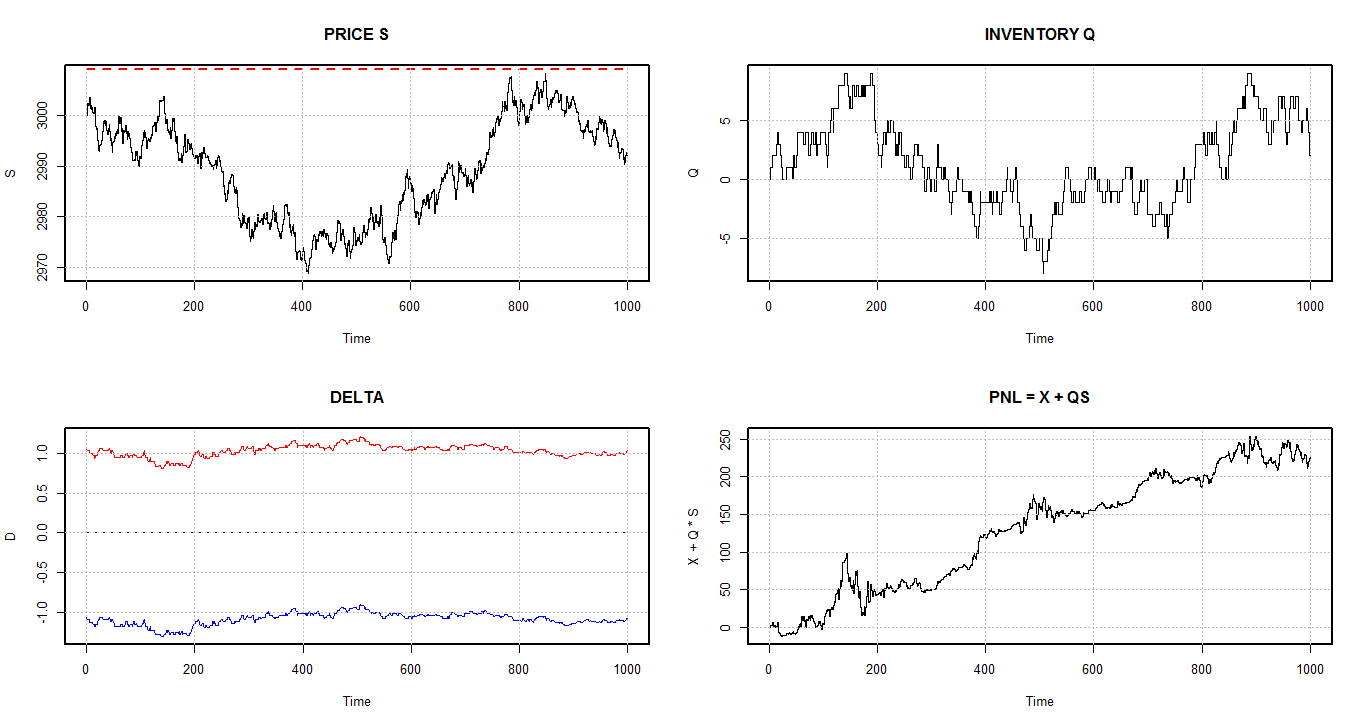}
\caption{A trading day with martingale controls.
\label{fig-day-M} }
\end{figure}

In Figure \ref{fig-day-M} we represent the martingale strategy, whose controls $\delta^\pm_\ast$ are rather constant, around $\pm 1.0$. They small fluctuations are due to inventory, but are independent of the mid-price. When the time $t<200$ the quotes are tilted downwards: the market-maker has a positive inventory, thus she improves her ask price in order to sell. At $t=200$ there is a jump upwards on the quotes, following a drop in the inventory. At the end of the day there is less fluctuation on the quotes: the inventory penalty due to the volatility is near zero, and the penalty at expiry does not trigger because the inventory is rather flat.\\

\begin{figure}
\centering
\includegraphics[width=5in]{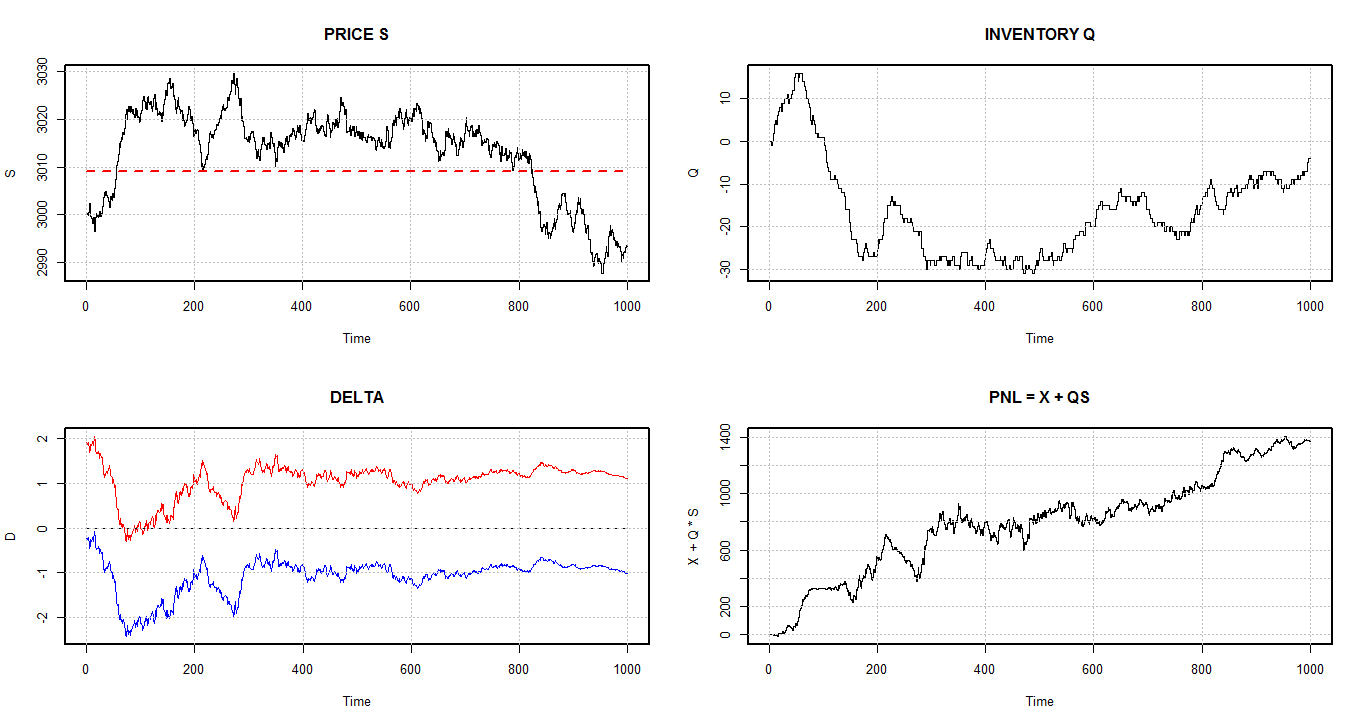}
\caption{A trading day with Ornstein-Uhlenbeck (i.e. mean reverting) controls.
\label{fig-day-OU} }
\end{figure}

In Figure \ref{fig-day-OU} the quotes have stronger fluctuations. Near $t=0$ the quotes are downward, reflecting the fact that the mid-price is below its mean $\mu$ (red-dotted line) and the market-maker bets the price will go up. Around $t=50$ the price crosses $\mu$ and sits above $\mu$: the market-maker bets for prices to come back later, hence she gets aggresive selling quotes and builds up a negative inventory. At $t=200$, when the price comes back to $\mu$, she starts to unwind some of her position. At the end of the day the mid-price plunges below $\mu$, but there market-maker does not follow the buying signal: the dynamic is no longer driven by a directional bet but rather on clearing her inventory, which explains the steady slope in her inventory towards zero instead of a peak upward.\\

\begin{figure}
\centering
\includegraphics[width=6in]{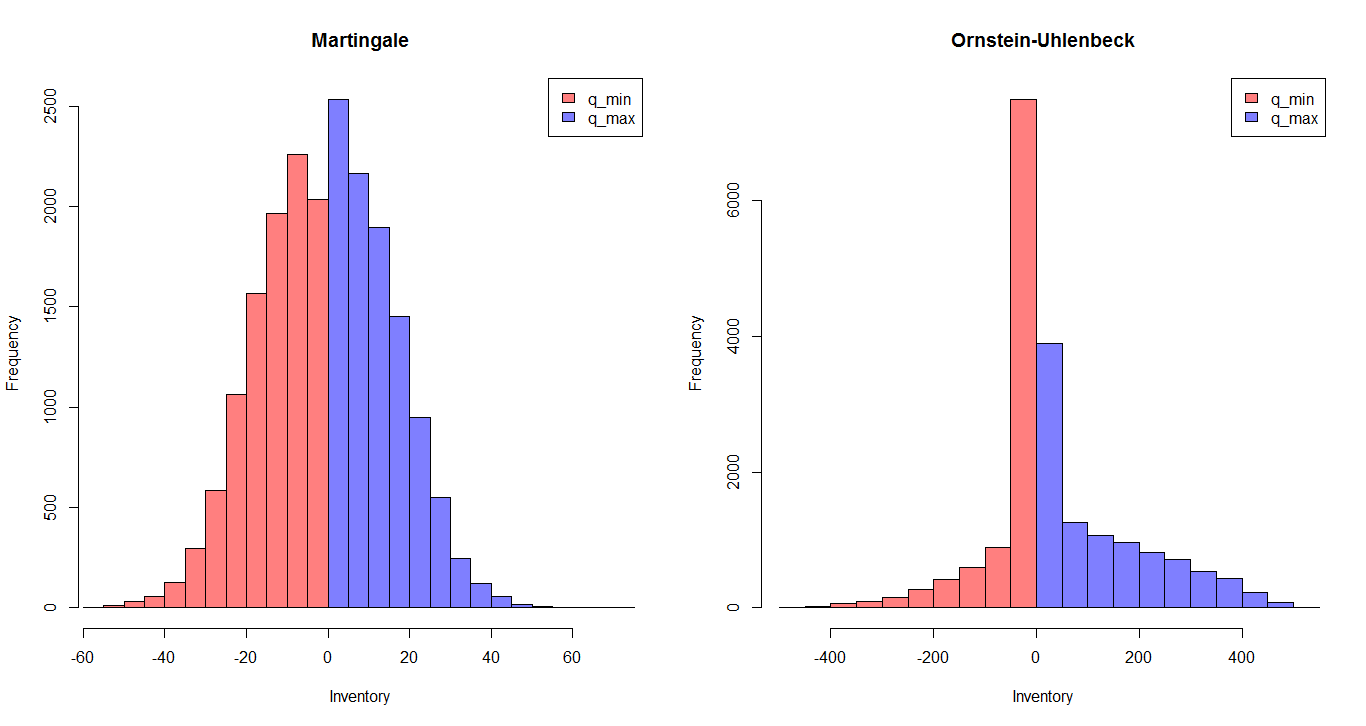}\\
\caption{ Min-Max intraday inventory.
\label{fig-histo-inv}}
\end{figure}

\begin{figure}
\centering
\includegraphics[width=6in]{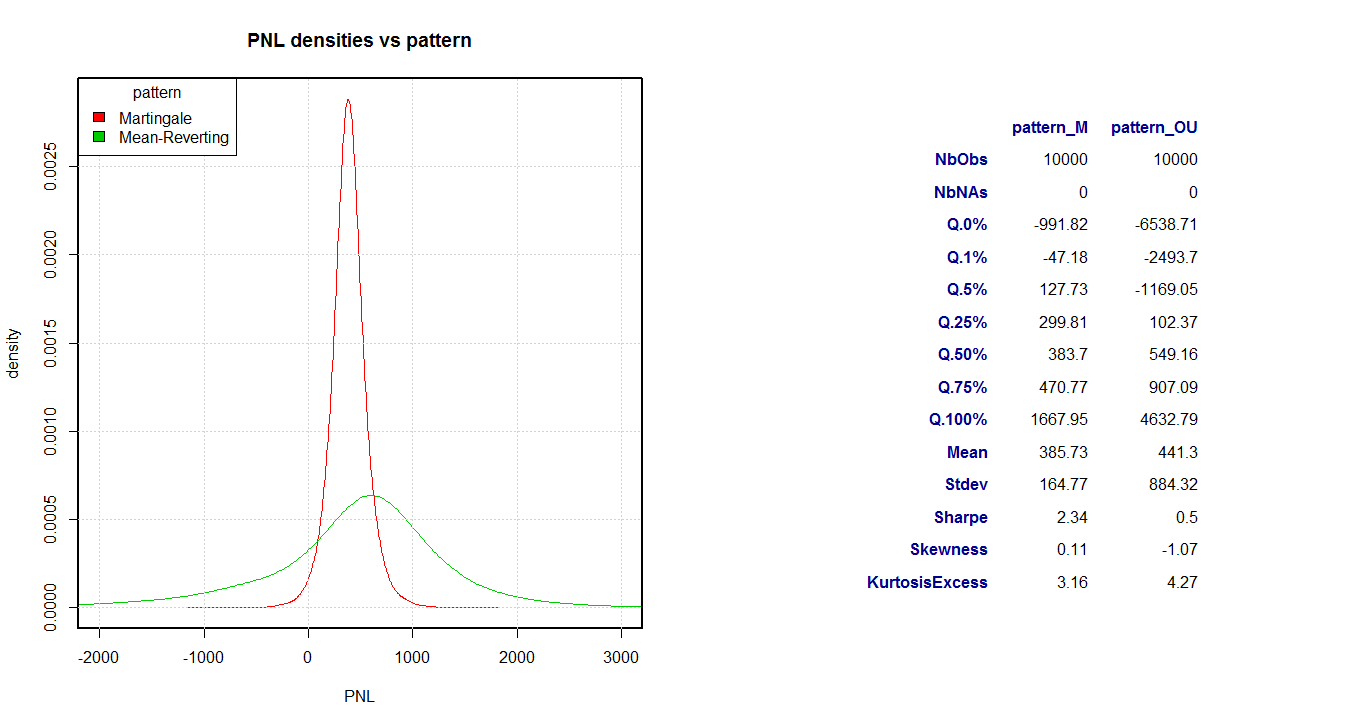}\\
\caption{Comparison of directional patterns: martingale (M) vs mean-reverting (OU).
\label{fig-density-policy}}
\end{figure}

\begin{figure}
\centering
\includegraphics[width=6in]{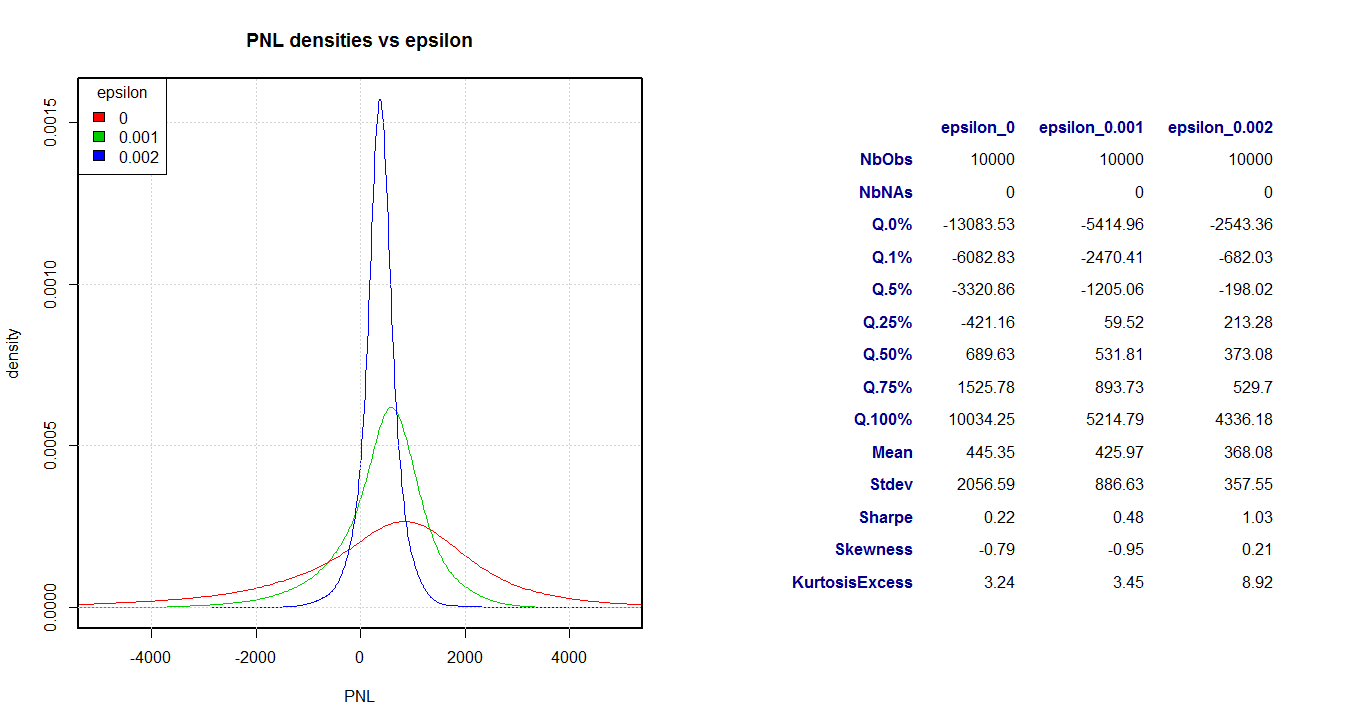}\\
\caption{The effect of the inventory risk $\varepsilon$ on the mean-reverting strategy.
\label{fig-density-epsilon}}
\end{figure}

\begin{figure}
\centering
\includegraphics[width=6in]{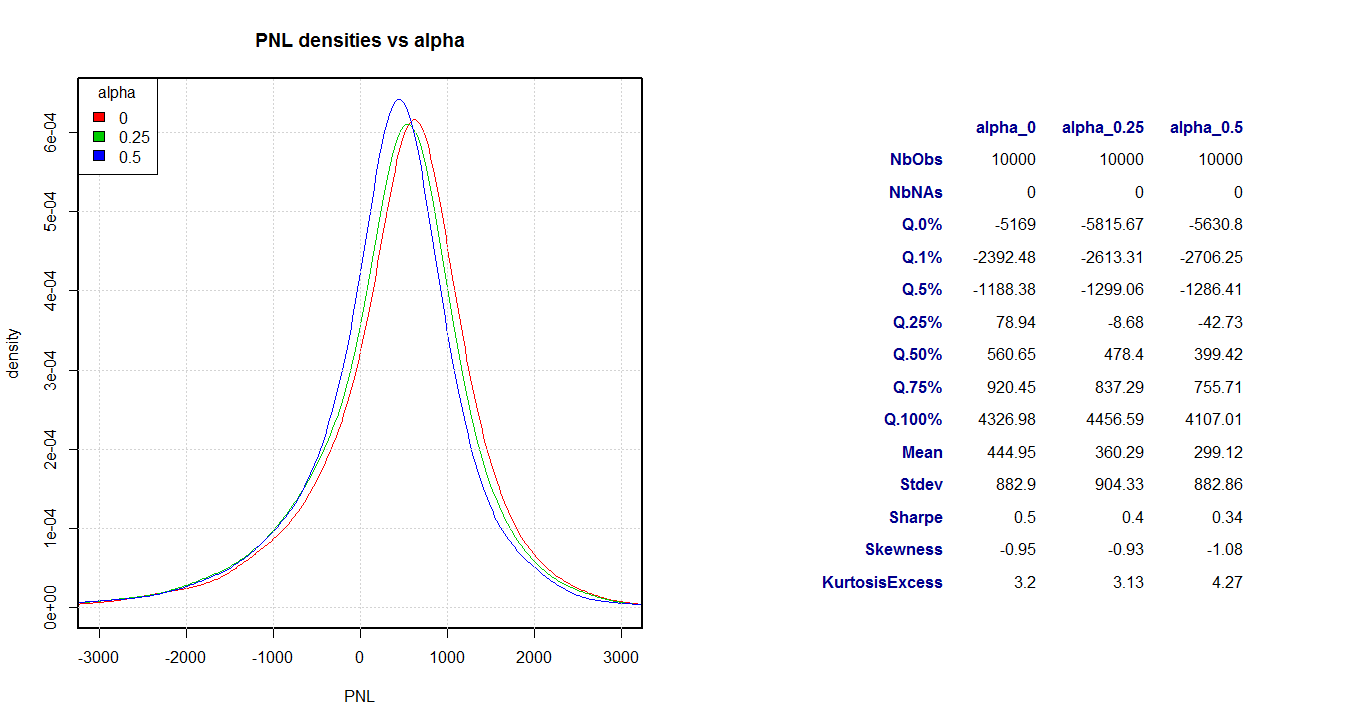}\\
\caption{The effect of transaction costs $\alpha$ on the mean-reverting strategy.
\label{fig-density-alpha}}
\end{figure}

In Figure \ref{fig-histo-inv} we plotted the histogram of the maximum and minimum intraday inventory. On the left-hand side we have the martingale strategy, which is rather symmetric: the strategy has no directional bet, although the mid-price has an upward trend. On the right-hand side we have the mean-reverting strategy, whose histogram is skewed to the right. The market-maker tends to be positive in inventory because she bets the price will go up. The inventory risk is much bigger for Ornstein-Uhlenbeck than for the martingale: the inventory of the latter is around $[-60,60]$ whilst that of the former is around $[-400,500]$.\\

In Figure \ref{fig-density-policy} we compare the effect of the directional bet on the PNL density. On the one hand, the mean-reverting strategy earns more than the martingale one: it has higher mean, median and higher quantiles. This happens because the market-maker knows the pattern and the parameters of the mid-price, and exploits this information. On the other hand, the mean-reverting strategy is more risky than the martingale one, as it can be inferred from the lower quantiles and the higher-order moments (standard deviation, skewness, kurtosis). In consequence, a direcional bet makes the strategy more profitable but more risky.\\

In Figure \ref{fig-density-epsilon} we show the effect of the inventory-risk parameter $\varepsilon$ on the PNL density. It is clear that as $\varepsilon$ increases, the PNL density has thinner tails and is more concentrated, which corresponds to a substantial risk reduction, as we can see in the quantiles and the higher-order moments. Although there is a decrease in the average daily gain, the (daily) Sharpe ratio is increasing in $\varepsilon$.\\

In Figure \ref{fig-density-alpha} we see the effect of the transaction costs $\alpha$ on the PNL density. There is a significant reduction on both the average daily gain and the (daily) Sharpe ratio, which implies the market-making becomes less profitable in the presence of higher transaction costs. For the other statistics, there is no significant change. One could thus be tempted thus to think that the effect of $\alpha$ on the PNL density is just a shift towards the left, but that is not exact because the shift is visible on the right tail and the head but it is negligeable on the left tail. In consequence, the transaction costs reduce the performance but they do not reduce the overall risk profile.


\section{Multi-asset model}

Assume now that we have $M$ different assets or mid-prices $S_1,\dots,S_M$. Our approach allows for general Markov processes, but for simplicity (and without loss of generality) let us suppose that $S(t)$ is an $M$-dimensional It\^o diffusion
\[
dS(t) = a(t,S)dt + \Sigma(t,S)dW(t)\,,
\]
where $a(t,S)\in\R^M$ is the drift and $dW(t)$ is an $M$-dimensional standard Brownian motion. The $M\times M$ matrix $\Sigma(t,S)$ is the square root of the (symmetric, positive-defined) variance-covariance matrix $\Lambda$, i.e. $\Lambda = \Sigma^\prime \Sigma$.\
The inventory $Q(t)$ is driven by the $M$-dimensional couple $(N^+(t),N^-(t))$, where the $i^{th}$ component $(N^+_i(t),N^-_i(t))$ is a Cox (or inhomogeneous Poisson) process having controlled intensity:
\begin{equation}\label{eq-lambda-m}
\left(\lambda^+_i(\delta^+_i),\lambda^-_i(\delta^-_i)\right):=\left( A_ie^{-k_i[z_i+\delta^+_i]},A_ie^{-k_i[z_i+\delta^-_i]}\right)
\end{equation}
In summary, the SDEs that govern the dynamics of the processes are
\begin{eqnarray}\label{eq-SDE-m}
dX(t) &=& [S(t)+\delta^+(t)]\cdot dN^+(t)-[S(t)+\delta^+(t)]\cdot dN^-(t)\,,\\
dS(t) &=& a(t,S)dt + \Sigma(t,S)dW(t),\nonumber\\
dQ(t) &=& -dN^+(t)+dN^-(t)\,, \nonumber
\end{eqnarray}
where $X(t)\in\mathbb{R}$, $Q(t)\in\mathbb{Z}^M$ and $S(t)\in\mathbb{R}^M$.

\subsection{Inventory penalties}

As before, we have two penalties.

\begin{enumerate}
\item A penalty at expiry,
\[
\Pi_1 := \eta Q^\prime(T)\Omega(T)Q(T)\,,\quad \eta\ge0\,,
\]
where $\Omega$ is an $M\times M$ symmetric, positive-definite matrix depending on some stochastic processes. For example, if the penalty is the cost of passing market orders at the end of the trading day then then $\Omega_{ij} = \delta_{ij}Z_i$, i.e. $\Omega$ is diagonal and coincides with the vector spread. However, $\Omega$ could be any matrix that takes into account the cost of executing the $M$-dimensional portfolio $Q(T)$ at market, including (but not limited to) the variance-covariance $\Lambda = \Sigma^\prime\Sigma$.

\item An integral (thus path-dependent) penalty
\[
\Pi_2 := \nu \int_t^T Q^\prime(\xi)\Lambda(\xi)Q(\xi)\,d\xi\,.
\]
\end{enumerate}

For each asset $S_i$ we have two controls $\delta^\pm_i$ in an admissible space $\boA_i$. this means that the vector controls $\delta^\pm = (\delta^\pm_1,\dots,\delta^\pm_M)$ lie in the admissible space $\boA = \boA_1\times\cdots\times\boA_m$. Under this framework, The value function is thus:
\begin{equation}\label{eq-u-z-m}
u(t,y,q,x) = \max_{\delta^\pm\in\boA}\E_{t,y,q,x}\Bigl[X(T)+Q(T)\cdot S(T)-\varepsilon \Pi(T)\Bigr]\,,\quad \Pi := \Pi_1 +\Pi_2\,,
\end{equation}

\subsection{The HJB equation and the ansatz}

The HJB equation is
\begin{eqnarray}\label{eq-hjb-epsi-m-gen}
\left(\partial_t+\boL\right)u + \max_{\delta^\pm\in\boA}\PP[\text{jump}]\left[u(t,y^+,q^+,x^+)-u(t,y,q,x)\right] &=& \varepsilon\nu q^\prime \Lambda q\,,\\
u(T,y,q,x) &=&x+s\cdot q-\varepsilon\eta q^\prime \Omega q \,,\nonumber
\end{eqnarray}
where $\PP[\text{jump}]$ is the probability of having a jump (i.e. market orders hitting the market-maker's quotes) at time $t$ and $u(t,y^+,q^+,x^+)$ is the value function after the jump. As it is customary with several independent Poisson jumps, we will assume that they cannot occur simultaneously. This means that the jump
\[
u(t,y^+,q^+,x^+)-u(t,y,q,x)
\]
can be decomposed as the sum of the jumps over $i$. More precisely, denote $\widehat{q_i}$ the vector $q$ without its $i$-th coordinate, i.e modulo index permutation we have $ q = (\widehat{q_i},q_i)$. Then the jumps are
\begin{eqnarray*}
u(t^+,y^+,q^+,x^+)-u(t,y,q,x) &=& \sum_{i=1}^M \Big[u(t,y,\widehat{q_i},q_i-1,x+(s_i+\delta^+_i))-u(t,y,q,x)\Big]\\
 &+& \sum_{i=1}^M \Big[u(t,y,\widehat{q_i}, q_i+1,x-(s_i-\delta^-_i))-u(t,y,q,x)\Big]\,.
\end{eqnarray*}
In consequence, the HJB equation \eqref{eq-hjb-epsi-m-gen} takes the form
\begin{eqnarray}\label{eq-hjb-epsi-m}
\left(\partial_t+\boL\right)u + \sum_{i=1}^M \max_{\delta^+_i\in\boA_i} Ae^{-k_i[z_i+\delta^+_i]}\Big[u(t,y,\widehat{q_i}, q_i-1,x+(s_i+\delta^+_i))-u(t,y,q,x)\Big] \\
 + \sum_{i=1}^M \max_{\delta^-_i\in\boA_i} Ae^{-k_i[z_i+\delta^-_i]}\Big[u(t,y,\widehat{q_i}, q_i+1,x-(s_i-\delta^-_i))-u(t,y,q,x)\Big] &=& \varepsilon\nu q^\prime \Lambda q\,,\nonumber\\
u(T,y,q,x) = x+s\cdot q &-&\varepsilon\eta q^\prime \Omega q \,.\nonumber
\end{eqnarray}

We will again use perturbation methods on $\varepsilon$ and search for a solution of \eqref{eq-hjb-epsi-m} with the following shape:
\begin{eqnarray}\label{eq-ans-z-m}
u(t,y,q,x) &=& x + v^{(0)}_q(t,y) + \varepsilon v^{(1)}_q(t,y) + O\left(\varepsilon^2\right)\,,\\
v^{(0)}_q(t,y) &=& \theta^{(0)}_0(t,y) + \sum_{i=1}^M q_i\theta^{(0)}_i(t,y)\,,\nonumber\\
v^{(1)}_q(t,y) &=& \theta^{(1)}_0(t,y) + \sum_{i=1}^M q_i\theta^{(1)}_i(t,y) + \sum_{i=1}^M\sum_{j=1}^M q_i q_j\theta^{(1)}_{ij}(t,y)\,,\nonumber
\end{eqnarray}
where $\theta^{(1)}_{ij}=\theta^{(1)}_{ji}$.

\subsection{Verification equation and its linearisation}

With the ansatz \eqref{eq-ans-z-m}, the marginal jumps can be readily computed:
\begin{eqnarray}\label{eq-jumps-m}
v^{(0)}_{q_i-1} - v^{(0)}_{q_i} &=& -\theta^{(0)}_i\,,\\
v^{(0)}_{q_i+1} - v^{(0)}_{q_i} &=& \theta^{(0)}_i\,,\nonumber\\
v^{(1)}_{q_i-1} - v^{(1)}_{q_i} &=& -\theta^{(1)}_i - 2\sum_{j=1}^M q_j\theta^{(1)}_{ij} + \theta^{(1)}_{ii}\,,\nonumber\\
v^{(1)}_{q_i+1} - v^{(1)}_{q_i} &=& \theta^{(1)}_i + 2\sum_{j=1}^M q_j\theta^{(1)}_{ij} + \theta^{(1)}_{ii}\,.\nonumber
\end{eqnarray}
Therefore, since the marginal optimal controls are
\begin{eqnarray*}
\delta^+_{i\ast} &=& \frac{1}{k_i}+s_i+v^{(0)}_{q_i}-v^{(0)}_{q_i-1} + \varepsilon\left[v^{(1)}_{q_i}-v^{(1)}_{q_i-1} \right] + O\left(\varepsilon^2\right)\,,\\
\delta^-_{i\ast} &=& \frac{1}{k_i}+s_i+v^{(0)}_{q_i}-v^{(0)}_{q_i+1} + \varepsilon\left[v^{(1)}_{q_i}-v^{(1)}_{q_i+1} \right] + O\left(\varepsilon^2\right)
\end{eqnarray*}
it follows that
\begin{eqnarray}\label{eq-delta-z-m}
\delta^+_{i\ast} &=& \frac{1}{k_i}-s_i+\theta^{(0)}_i + \varepsilon\left[\theta^{(1)}_i + 2\sum_{j=1}^M q_j\theta^{(1)}_{ij} - \theta^{(1)}_{ii}\right] + O\left(\varepsilon^2\right)\,,\\
\delta^-_{i\ast} &=& \frac{1}{k_i}+s - \theta^{(0)}_i + \varepsilon\left[-\theta^{(1)}_i - 2\sum_{j=1}^M q_j\theta^{(1)}_{ij} - \theta^{(1)}_{ii} \right] + O\left(\varepsilon^2\right)\,.\nonumber
\end{eqnarray}
Under these conditions, the verification equation is
\begin{eqnarray}\label{eq-verif-nonlin1-m}
\left(\partial_t+\boL\right)\left(v^{(0)}_q + \varepsilon v^{(1)}_q\right) &+& \sum_{i=1}^M\frac{A_i}{ek_i}e^{-k_iz_i}e^{k_is_i}e^{-k_i\theta^{(0)}_i} \exp\left\{ -k_i\varepsilon\left[\theta^{(1)}_i + 2\sum_{j=1}^M q_j\theta^{(1)}_{ij} - \theta^{(1)}_{ii} \right] + O(\varepsilon^2)\right\} \nonumber\\
&+& \sum_{i=1}^M\frac{A_i}{ek_i}e^{-k_iz_i}e^{-k_is_i}e^{k_i\theta^{(0)}_i} \exp\left\{ k_i\varepsilon\left[\theta^{(1)}_i + 2\sum_{j=1}^M q_j\theta^{(1)}_{ij} + \theta^{(1)}_{ii} \right] + O(\varepsilon^2)\right\}\nonumber\\
&=& \varepsilon\nu q^\prime\Lambda q\,,\nonumber\\
v^{(0)}_q(T,y) &=& s\cdot q\,,\\
v^{(1)}_q(T,y) &=& -\eta q^\prime \Omega q\,.\nonumber
\end{eqnarray}
The linearisation in $\varepsilon$ of the verification equation \eqref{eq-verif-nonlin1-m} is
\begin{eqnarray}\label{eq-verif-lin1-m}
\left(\partial_t+\boL\right)\left(v^{(0)}_q + \varepsilon v^{(1)}_q\right) &+& \sum_{i=1}^M\frac{A_i}{ek_i}e^{-k_iz_i}e^{k_is_i}e^{-k_i\theta^{(0)}_i} \left\{1 -k_i\varepsilon\left[\theta^{(1)}_i + 2\sum_{j=1}^M q_j\theta^{(1)}_{ij} - \theta^{(1)}_{ii} \right] + O(\varepsilon^2)\right\} \nonumber\\
&+& \sum_{i=1}^M\frac{A_i}{ek_i}e^{-k_iz_i}e^{-k_is_i}e^{k_i\theta^{(0)}_i}\left\{1+ k_i\varepsilon\left[\theta^{(1)}_i + 2\sum_{j=1}^M q_j\theta^{(1)}_{ij} + \theta^{(1)}_{ii} \right] + O(\varepsilon^2)\right\}\nonumber\\
&=& \varepsilon\nu q^\prime\Lambda q\,,\nonumber\\
v^{(0)}_q(T,y) &=& s\cdot q\,, \\
v^{(1)}_q(T,y) &=& -\eta q^\prime \Omega q\,.\nonumber
\end{eqnarray}

\subsection{Solution for $\varepsilon^0$}

At zero-th order in $\varepsilon$ we get
\begin{eqnarray}\label{eq-verif-epsi0-m}
\left(\partial_t+\boL\right)v^{(0)}_q  + \sum_{i=1}^M\frac{2A_i}{ek_i}e^{-k_iz_i}\cosh\left(k_i[\theta^{(0)}_i-s_i]\right) &=& 0\,,\\
v^{(0)}_q(T,y) &=& s\cdot q\,. \nonumber
\end{eqnarray}
We decompose \eqref{eq-verif-epsi0-m} in its $M+1$ components. For $\theta^{(0)}_0$ we obtain
\begin{eqnarray}\label{eq-verif-epsi0-0-m}
\left(\partial_t+\boL\right)\theta^{(0)}_0  + \sum_{i=1}^M\frac{2A_i}{ek_i}e^{-k_iz_i}\cosh\left(k_i[\theta^{(0)}_i-s_i]\right) &=& 0\,,\\
\theta^{(0)}_0(T,y) &=& 0\,, \nonumber
\end{eqnarray}
whilst for $\theta^{(0)}_i$ we get
\begin{eqnarray}\label{eq-verif-epsi0-i-m}
\left(\partial_t+\boL\right)\theta^{(0)}_i &=& 0\,,\\
\theta^{(0)}_i(T,y) &=& s_i\,. \nonumber
\end{eqnarray}
Using Feynman-Kac (recursively) we obtain
\begin{eqnarray}\label{eq-theta-epsi0-m}
\theta^{(0)}_i(t,y)  &=& s_i+\Delta_i\,,\\
\theta^{(0)}_0(t,y)  &=& \sum_{i=1}^M\frac{2}{e}\mathbb{E}_{t,y}\left[\int_t^T\frac{A_i}{K_i}e^{-K_iZ_i}\cosh(K_i\Delta_i)\,d\xi\right]\,,\nonumber
\end{eqnarray}
where $\Delta_i := \mathbb{E}_{t,y}\left[S_i(T)\right] - s_i$. As in the single-asset case, the solution 
\[
u(t,y,q,x) = x + \sum_{i=1}^M q_i(s_i+\Delta_i) + 
\sum_{i=1}^M\frac{2}{e}\mathbb{E}_{t,y}\left[\int_t^T\frac{A_i}{K_i}e^{-K_iZ_i}\cosh(K_i\Delta_i)\,d\xi\right]
\]
is exact for $\varepsilon = 0$, i.e. for the problem without inventory constraints. As before, the first two terms of the right-hand side are the buy-and-hold strategy, i.e. the expected gain of keeping the current inventory position until the end of the trading day. The third term takes into account the market-making thoughout the day (that is why it has an integral term) with a directional bet (given by the $\Delta_i$'s). Since the third term is strictly positive then it is more profitable to play market-making than buy-and-hold only, and it works better for non-martingale mid-price processes.

\subsection{Solution for $\varepsilon^1$}

At first order in $\varepsilon$ we have
\begin{eqnarray}\label{eq-verif-epsi1-m}
\left(\partial_t+\boL\right) v^{(1)}_q &-& \sum_{i=1}^M\frac{A_i}{e}e^{-k_iz_i}e^{-k_i\Delta_i} \left[\theta^{(1)}_i + 2\sum_{j=1}^M q_j\theta^{(1)}_{ij} - \theta^{(1)}_{ii} \right] \nonumber\\
&+& \sum_{i=1}^M\frac{A_i}{e}e^{-k_iz_i}e^{k_i\Delta_i}\left[\theta^{(1)}_i + 2\sum_{j=1}^M q_j\theta^{(1)}_{ij} + \theta^{(1)}_{ii} \right] \nonumber\\
&=& \nu q^\prime\Lambda q\,,\\
v^{(1)}_q(T,y) &=& -\eta q^\prime \Omega q\,.\nonumber
\end{eqnarray}
We decompose \eqref{eq-verif-epsi1-m} into several components. For $\theta^{(1)}_0$ we obtain
\begin{eqnarray}\label{eq-verif-epsi1-0-m}
\left(\partial_t+\boL\right)\theta^{(1)}_0  + \sum_{i=1}^M\frac{2A_i}{ek_i}e^{-k_iz_i}\sinh\left(k_i\Delta_i\right)\theta^{(1)}_i 
 + \sum_{i=1}^M\frac{2A_i}{ek_i}e^{-k_iz_i}\cosh\left(k_i\Delta_i\right)\theta^{(1)}_{ii} &=& 0\,,\\
\theta^{(1)}_0(T,y) &=& 0\,; \nonumber
\end{eqnarray}
for $\theta^{(1)}_i$ we have
\begin{eqnarray}\label{eq-verif-epsi1-1-m}
\left(\partial_t+\boL\right)\theta^{(1)}_i + \frac{4A_i}{e}e^{-k_iz_i}\sinh(k_i\Delta_i)\sum_{j=1}^M\theta^{(1)}_{ij}&=& 0\,.\nonumber\\
\theta^{(1)}_i(T,y) &=& 0\,;
\end{eqnarray}
and for $\theta^{(1)}_{ij}$ we get
\begin{eqnarray}\label{eq-verif-epsi1-2-m}
\left(\partial_t+\boL\right)\theta^{(1)}_{ij} &=& \nu\Lambda_{ij}\,,\\
\theta^{(1)}_{ij}(T,y) &=& -\eta\Omega_{ij}\,.\nonumber
\end{eqnarray}
Solving (recursively) with Feynmann-Kac yields
\begin{eqnarray}\label{eq-theta-epsi1-m}
\theta^{(1)}_{ij}(t,y) &=& -\tilde\pi_{ij}\,,\\
\theta^{(1)}_i(t,y) &=& -\frac{4}{e}\E_{t,y}\left[\int_t^T A_ie^{-K_iZ_i} \sinh(K_i\Delta_i)\sum_{j=1}^M\tilde\pi_{ij}\,d\xi\right]\,,\nonumber\\
\theta^{(1)}_0(t,y) &=& \frac{2}{e}\sum_{j=1}^M\E_{t,y}\left[\int_t^T A_ie^{-K_iZ_i}\left\{\theta^{(1)}_i\sinh(K_i\Delta_i)-\tilde\pi_{ij}\cosh(K_i\Delta_i)\right\}\,d\xi\right]\,,\nonumber
\end{eqnarray}
where
\[
\tilde\pi_{ij} := \eta\E_{t,y}\left[\Omega_{ij}(T)\right]+
\nu\E_{t,y}\left[\int_t^T\Lambda_{ij}\,d\xi\right]\,.
\]
Putting all together \eqref{eq-ans-z-m}, \eqref{eq-theta-epsi0-m}, \eqref{eq-theta-epsi1-m} we obtain the explicit expansion of $u(t,y,q,x)$ up to first order in $\varepsilon$.

\subsection{Optimal controls}

Plugging the explicit expressions \eqref{eq-theta-epsi0-m}, \eqref{eq-theta-epsi1-m} into \eqref{eq-delta-z-m} yields
\begin{eqnarray}\label{eq-delta-i-m}
\delta^+_{i\ast} &=& \frac{1}{k_i}+\Delta_i + \varepsilon\left\{-\frac{4}{e}\E_{t,y}\left[\int_t^T A_ie^{-K_iZ_i} \sinh(K_i\Delta_i)\sum_{j=1}^M\tilde\pi_{ij}\,d\xi\right] - 2\sum_{j=1}^M q_j\tilde\pi_{ij} + \tilde\pi_{ii}\right\} + O\left(\varepsilon^2\right)\,,\nonumber\\
 & & \\
\delta^-_{i\ast} &=& \frac{1}{k_i}-\Delta_i + \varepsilon\left\{+\frac{4}{e}\E_{t,y}\left[\int_t^T A_ie^{-K_iZ_i}\sinh(K_i\Delta_i)\sum_{j=1}^M\tilde\pi_{ij}\,d\xi\right\} + 2\sum_{j=1}^M q_j\tilde\pi_{ij} + \tilde\pi_{ii} \right] + O\left(\varepsilon^2\right)\,.\nonumber
\end{eqnarray}
Let us lighten the notation via vectors. If we define
\[
\delta^\pm_\ast = \left[\begin{array}{c}
\delta^\pm_{1\ast}\\
\vdots\\
\delta^\pm_{M\ast}
\end{array}\right] \,,\quad k^{-1}=\left[\begin{array}{c}
1/k_1\\
\vdots\\
1/k_M
\end{array}\right]\,,\quad \Delta=\left[\begin{array}{c}
\Delta_1\\
\vdots\\
\Delta_M
\end{array}\right]\,,\quad \mathbbm{1}=\left[\begin{array}{c}
1\\
\vdots\\
1
\end{array}\right]\,,
\]
\[
\tilde\pi=\left[\begin{array}{ccc}
\tilde\pi_{11} & \cdots & \tilde\pi_{1M}\\
\vdots &  & \vdots\\\
\tilde\pi_{M1} & \cdots & \tilde\pi_{MM}
\end{array}\right]\,,\quad \mathrm{Diag}(\tilde\pi)=\left[\begin{array}{c}
\tilde\pi_{11}\\
\vdots\\\
\tilde\pi_{MM}
\end{array}\right]\,,
\]
and
\[
H=\frac{4}{e}\left[\begin{array}{ccc}
A_ie^{-K_1Z_1}\sinh(K_1\Delta_1) &  & 0\\
 & \ddots & \\\
0 &  & A_Me^{-K_MZ_M}\sinh(K_M\Delta_M)
\end{array}\right]\,,
\]
then the optimal controls are
\begin{eqnarray}\label{eq-delta-vec-m}
\delta^+_{\ast} &=& k^{-1}+\Delta + \varepsilon\left\{-\E_{t,y}\left[\int_t^T H\tilde\pi\mathbbm{1}\,d\xi\right] - 2\tilde\pi q + \mathrm{Diag}(\tilde\pi)\right\} + O\left(\varepsilon^2\right)\,,\\
\delta^-_{\ast} &=& k^{-1}-\Delta + \varepsilon\left\{+\E_{t,y}\left[\int_t^T H\tilde\pi\mathbbm{1}\,d\xi\right] + 2\tilde\pi q + \mathrm{Diag}(\tilde\pi)\right\} + O\left(\varepsilon^2\right)\,,\nonumber\\
\psi_{\ast} &=& 2k^{-1} + 2\varepsilon \mathrm{Diag}(\tilde\pi)+ O\left(\varepsilon^2\right)\,,\nonumber\\
r_{\ast} &=& s + \Delta - \varepsilon\left\{\E_{t,y}\left[\int_t^T H\tilde\pi\mathbbm{1}\,d\xi\right] + 2\tilde\pi q\right\} + O\left(\varepsilon^2\right)\,.\nonumber
\end{eqnarray}

\subsection{The effect of transaction costs}

For each asset $S_i$ we can add a transaction cost $\alpha_i$. In order to use the vectorial notation as in \eqref{eq-delta-vec-m}, define
\[
\alpha=\left[\begin{array}{c}
\alpha_1\\
\vdots\\
\alpha_M
\end{array}\right]
\]
and
\[
H(\alpha)=\frac{4}{e}\left[\begin{array}{ccc}
A_ie^{-K_1(Z_1+\alpha_1)}\sinh(K_1\Delta_1) &  & 0\\
 & \ddots & \\\
0 &  & A_Me^{-K_M(Z_M+\alpha_M)}\sinh(K_M\Delta_M)
\end{array}\right]\,.
\]
It can be shown that the optimal controls under transaction costs are
\begin{eqnarray}\label{eq-delta-vec-alpha}
\delta^+_{\alpha\ast} &=& k^{-1}+ \alpha + \Delta + \varepsilon\left\{-\E_{t,y}\left[\int_t^T H(\alpha)\tilde\pi\mathbbm{1}\,d\xi\right] - 2\tilde\pi q + \mathrm{Diag}(\tilde\pi)\right\} + O\left(\varepsilon^2\right)\,,\\
\delta^-_{\alpha\ast} &=& k^{-1} + \alpha - \Delta + \varepsilon\left\{+\E_{t,y}\left[\int_t^T H(\alpha)\tilde\pi\mathbbm{1}\,d\xi\right] + 2\tilde\pi q + \mathrm{Diag}(\tilde\pi)\right\} + O\left(\varepsilon^2\right)\,.\nonumber\\
\psi_{\alpha\ast} &=& 2k^{-1} + 2\alpha + 2\varepsilon \mathrm{Diag}(\tilde\pi)+ O\left(\varepsilon^2\right)\,,\nonumber\\
r_{\alpha\ast} &=& s + \Delta - \varepsilon\left\{\E_{t,y}\left[\int_t^T H(\alpha)\tilde\pi\mathbbm{1}\,d\xi\right] + 2\tilde\pi q\right\} + O\left(\varepsilon^2\right)\,.\nonumber
\end{eqnarray}
Observe that the optimal market-maker's spread $\psi_{\alpha\ast}$ do not have crossed terms, i.e. the spreads are independent. The optimal centres $r_{\alpha\ast}$ do have cross-correlation terms involved via $\tilde\pi$, which enter into play twice: first, as weights on the directional bet $H(\alpha)$, integrated over the remaining time $T-t$; second, as weights on the inventory $q$.\\

In the particular case where all mid-prices are martingales, i.e. $\Delta=0$ and $H(\alpha)=0$, we have 
\begin{eqnarray}\label{eq-delta-vec-mart}
\psi_{\alpha\ast} &=& 2k^{-1} + 2\alpha + 2\varepsilon \mathrm{Diag}(\tilde\pi)+ O\left(\varepsilon^2\right)\,,\\
r_{\alpha\ast} &=& s - 2\varepsilon \tilde\pi q + O\left(\varepsilon^2\right)\,.\nonumber
\end{eqnarray}

\subsection{Risk management and iso-risk surfaces}

Recall that the inventory risk is given by $\Pi(q)=\varepsilon[\Pi_1(q)+\Pi_2(q)]$, where
\[
\Pi_1(q) = \eta q^\prime\Omega q\,,\quad \Pi_2(q) := \nu \int_t^T q^\prime\Lambda q\,d\xi
\]
with both $\Omega$ and $\Lambda$ are $M\times M$ symmetric, positive-definite matrices. The $(M-1)$-dimensional iso-risk surface at level $c\ge0$ is defined as
\[
\mathrm{isoRisk}(c) =\left\{w\in\R^M: \Pi(w)=c\right\}\,.
\]
Observe that $q^\prime\Omega q$ and $q^\prime\Lambda q$ are positive-definite bilinear forms in $\R^M$, hence they define two norms  equivalent to the Euclidean one. Moreover, their corresponding level surfaces are ellipsoids. This implies that the iso-risk surface $\mathrm{isoRisk}(c)$ is determined by the level curves of $\Pi_1$ and $\Pi_2$, i.e. by the norms of the inventory vector $q$ induced by $\Omega$ and $\Lambda$.\\

In order to assess the effect of cross-correlation in the risk measure, let us consider the simplest example, i.e. $M=2$, $\varepsilon=1$, $\eta=1$, $\nu=0$ and
\[
\Omega = \left[\begin{array}{cc}
1 & \rho\\
\rho & 1
\end{array}\right]\,,\quad\rho\in[0,1)\,.
\]
In this case, the iso-risk curves are ellipses, i.e.
\[
\mathrm{isoRisk}(c) =\left\{w=(w_1,w_2)^\prime\in\R^2: \Pi(w) = c\right\}\,,\quad \Pi(w) = w_1^2 + 2\rho w_1w_2 + w_2^2\,.
\]
Now, assume that the market-maker looks for an inventory $q=(q_1,q_2)^\prime$ such that $\vert q_1\vert + \vert q_1\vert = 2$. Then the possible configurations (modulo the symmetry $q_1\leftrightarrow q_2)$ are
\[
q_A=(2,0)^\prime\,,\quad q_B=(1,1)^\prime\quad\text{and}\quad q_C=(1,-1)^\prime\,.
\]
Since
\[
\Pi(q_A)=4\,,\quad \Pi(q_B)=2(1+\rho)\,,\quad \Pi(q_C)=2(1-\rho)\,,
\]
if $\rho\in(0,1)$ then
\[
\Pi(q_C)<\Pi(q_B)<\Pi(q_A)\,,
\]
i.e. all three configurations belong to a different iso-risk ellipse and $q_C$ is the unique configuration minimising the risk. Therefore, the market-maker will privilege $q_C$ over $q_A$ and $q_B$.\\

Imagine that the current inventory of the market-maker is $(2,-1)^\prime$, i.e. an absolute inventory of 3. If she desires to reduce her absolute inventory to 2 then she will choose $(1,-1)^\prime$ over $(2,0)^\prime$. However, it is not always desirable to reduce the absolute inventory: if $\rho>1/4$ then
\[
\Pi((2,-1)^\prime) = 5-4\rho<4= \Pi((2,0)^\prime)\,,
\]
which implies that the market-maker would prefer $(2,-1)^\prime$ over $(2,0)^\prime$, even if in terms of absolute inventory this choice seems counter-intuitive.\\

In the general case, the philosophy is the same: the risks in inventory are not measured in absolute terms, but in terms of the bilinear forms $q^\prime\Omega q$ and $q^\prime\Lambda q$ and how they affect the iso-risk surface $\mathrm{isoRisk}(c)$. Therefore, in the presence of non-zero correlations the market-maker will favour a well-diversified portfolio: the optimal inventory vector $q$ is computed following the same optimisation as for the Markowitz Portfolio and the Efficient Frontier.

\end{document}